\documentclass[11pt]{article}

\usepackage{graphicx}%
\usepackage{multirow}%
\usepackage{amsmath,amssymb,amsfonts}%
\usepackage{amsthm}%
\usepackage{mathrsfs}%
\usepackage[title]{appendix}%
\usepackage{xcolor}%
\usepackage{color}%
\usepackage{textcomp}%
\usepackage{manyfoot}%
\usepackage{booktabs}%
\usepackage{algorithm}%
\usepackage{algorithmicx}%
\usepackage{algpseudocode}%
\usepackage{listings}%
\usepackage{caption}
\usepackage{cleveref}
\usepackage{subcaption}
\usepackage{blindtext}
\usepackage[margin=2cm]{geometry}

\def \s{\mathbf{s}}
\def \v{\mathbf{v}}
\def \x{\mathbf{x}}
\def \r{\mathbf{r}}

\def \cm{\mathrm{cm}}

\def \sec{\mathrm{s}}
\def \K{\mathrm{K}}

\title{Cross-component energy transfer in superfluid helium-4}
\author{Piotr Z. Stasiak\textsuperscript{1}\footnote{p.stasiak@newcastle.ac.uk}, Andrew W. Baggaley\textsuperscript{1}, Giorgio Krstulovic\textsuperscript{2},\\ Carlo F. Barenghi\textsuperscript{1}, Luca Galantucci\textsuperscript{1,3}}
\date{}
\begin{document}
\maketitle

\vspace{-1cm}
\begin{center}
\textsuperscript{1}School of Mathematics, Statistics and Physics, Newcastle University,\\ Newcastle Upon Tyne, NE1 7RU, United Kingdom.\\
\textsuperscript{2}Universit\'{e} C\^ote d'Azur, Observatoire de la C\^ote d'Azur, CNRS,\\ Laboratoire Lagrange, Boulevard de l'Observatoire CS 34229 - F 06304\\ NICE Cedex 4, France.\\
\textsuperscript{3}Istituto per le Applicazioni del Calcolo M. Picone, IAC-CNR, Via dei \\Taurini 19, Roma, 00185, Italy.
\end{center}

\abstract{The reciprocal energy and enstrophy transfers 
between normal fluid and superfluid 
components dictate the overall dynamics of 
superfluid \textsuperscript{4}He including
the generation, evolution and coupling of coherent structures,
the distribution of energy among lengthscales,
and the decay of turbulence.
To better understand the essential ingredients of this
interaction, we employ a numerical
two-way model which self-consistently accounts for
the back-reaction of the superfluid vortex lines onto the normal fluid.
Here we focus on a prototypical laminar (non-turbulent) vortex configuration
which is simple enough to clearly relate 
the geometry of the vortex line to 
energy injection and dissipation to/from the normal fluid:
a Kelvin wave excitation on two vortex anti-vortex pairs evolving in (a) 
an initially quiescent normal fluid, and (b) an imposed counterflow. 
In (a), the superfluid injects energy and vorticity in the normal fluid.
In (b), the superfluid gains energy from the normal fluid 
via the Donnelly-Glaberson instability. 
}

\section{Introduction}\label{intro}

\begin{figure}[t]
\centering
\includegraphics[width=0.8\textwidth]{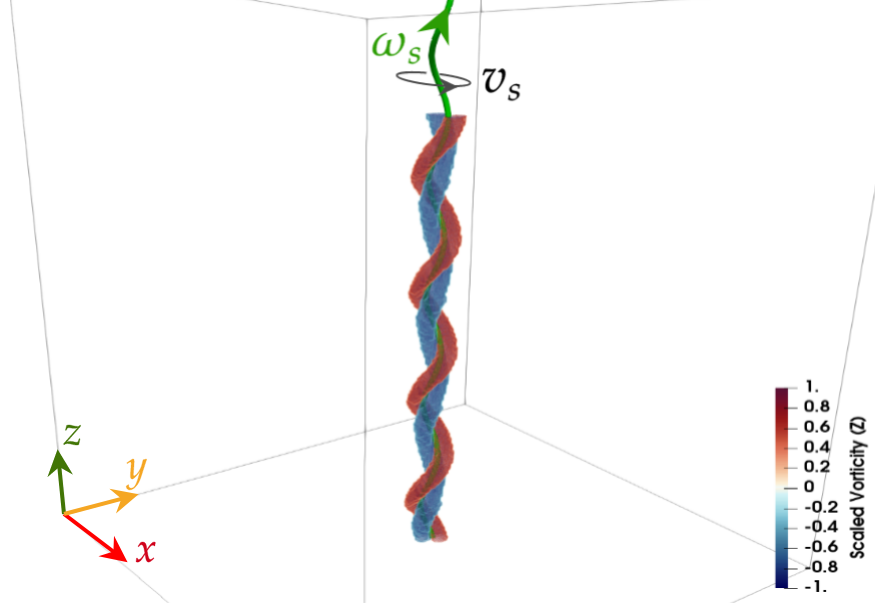}
\vspace{1ex}
\caption{\small
Schematic diagram of a single helical superfluid vortex (green) and the normal fluid vorticity in the $z$-direction (red and blue
for positive and negative vorticity values respectively).
The normal fluid vorticity
$\omega_z$ is normalised by the maximum value $\omega_{max}$. Note how the 
normal fluid structures wrap around the superfluid 
vortex line as a double-helix (the 3D realisation of the normal fluid dipole 
described in the literature).  The thickness of the superfluid
vortex core (green) is exaggerated for visual purposes; in reality it is 
several orders of magnitude smaller than the normal fluid's structures
shown in red and blue.
}
\label{fig: schematic}
\end{figure}

The turbulent flow of  helium~II (the low temperature phase of
liquid $^4$He) consists of a disordered tangle of interacting
superfluid vortex lines
which move in a thermal background of elementary excitations
\cite{vinen-niemela-2002,skrbek-sreenivasan-2012,
barenghi-skrbek-sreenivasan-2014, barenghi-etal-2023}. 
The vortex lines are topological defects of the  
superfluid component \cite{feynman-1955, onsager-1949,hall-vinen-1956a},
which is associated with the ground state, while the thermal
excitations constitute a viscous fluid (the normal
fluid component).  According to the two-fluid theory
\cite{tisza-1938,landau-1941},
normal fluid and superfluid are inseparable, penetrate each other
and comprise of independent density and velocity fields. 
Importantly, since the vortex lines act as scattering centres 
for the elementary excitations, normal fluid and superfluid 
are dynamically coupled \cite{hall-vinen-1956a}. This interaction,
called the
mutual friction force, transfers kinetic energy,
enstrophy and helicity 
\cite{kivotides-barenghi-samuels-2000,idowu-willis-barenghi-samuels-2000,
kivotides-2005} between the two fluids. 
The mutual friction therefore controls the generation of
normal fluid and superfluid coherent structures 
\cite{kivotides-barenghi-samuels-2000,kivotides-2015epl,kivotides2018superfluid,galantucci2023friction,peretti-et-al-2023} (see Fig. \ref{fig: schematic}) , 
their possible coupling \cite{morris-koplik-rouson-2008,kivotides-2011}, 
the spectral distribution of turbulent kinetic energy 
\cite{kivotides-2011,kivotides-2014}, as well as
the decay \cite{kivotides-2015jltp} and the dissipation 
of turbulent kinetic energy \cite{galantucci-etal-2023}. It is thus
responsible for some of the observed
similarities and differences between classical and quantum turbulence 
\cite{vinen-niemela-2002,skrbek-sreenivasan-2012,barenghi-skrbek-sreenivasan-2014,barenghi-etal-2023}. 

\begin{figure}
	\centering
	\includegraphics[width=0.9\textwidth]{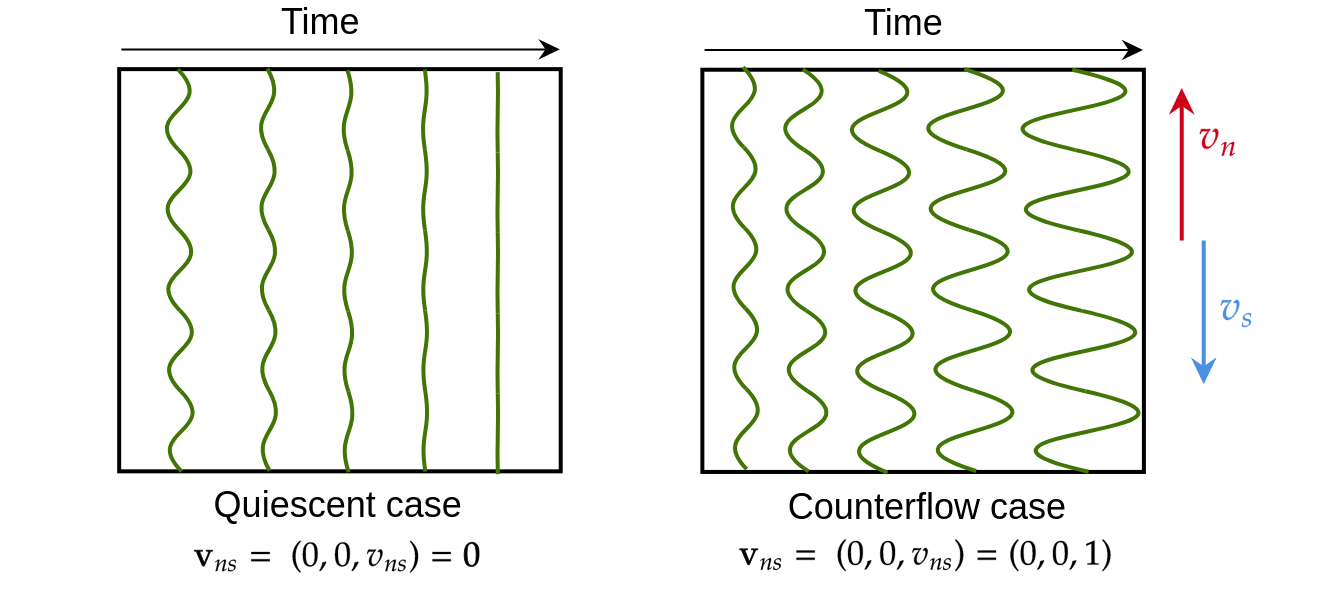}
	\caption{\small Schematic time evolution of the Kelvin wave
in the first problem (left: Kelvin wave in
initially quiescent normal fluid) and
in the second problem (right: Kelvin wave in the presence of
counterflow).}
\label{fig: 2d-schematic}
\end{figure}

Our work focuses on this energy transfer between the two 
fluids. In all previous studies, this topic was investigated
in the context of turbulence. In some studies
the normal fluid was initially  
turbulent and drove the growth of a superfluid vortex tangle 
\cite{kivotides-2006,kivotides-2011,kivotides-2014,kivotides-2015jltp});
in others, a superfluid vortex tangle excited an initially quiescent
normal fluid \cite{kivotides-2005,kivotides-2007}. 
Unfortunately, the turbulent context, requiring a statistical 
interpretation, complicates the analysis of the results.
In  order to better capture the essential physics of the
energy transfer, here we study  
the simpler dynamics (laminar rather than turbulent)
of two vortex-antivortex pairs, with each vortex perturbed by a single, helical Kelvin wave. We consider two problems (see Fig. \ref{fig: 2d-schematic}).
In the first problem, the vortex lines with Kelvin waves are embedded in
an initially quiescent normal fluid; in the second problem, they are in the presence of an
imposed counterflow (a configuration which
leads to the well-known Donnelly-Glaberson instability \cite{glaberson1974instability}).
Similar vortex configurations, currently examined in ongoing experiments 
\cite{peretti-et-al-2023}, are ubiquitous in turbulent flows, but they
are simple enough that we
can relate the temporal evolution of 
the geometry of the vortex line with the injection and the dissipation
of energy in the normal fluid via the mutual friction force.

The study utilises a recent theoretical
model capable of simulating self-consistently the coupled motion 
of normal fluid and superfluid \cite{galantucci2020new}; in particular
the model predicts quantitatively the 
experimentally observed shrinking of superfluid vortex rings 
\cite{tang2023imaging}. 

The paper is organised as follows: Section \ref{method} briefly
describes the model and the parameters used in our simulations; 
Section \ref{results} presents the results; Section \ref{conclusions} 
summarises the conclusions.

\section{Method}\label{method}

The vortex core radius of superfluid $^4$He ($a_0\approx10^{-8}\cm$) 
is several orders of magnitudes smaller
than any length scale of interest in turbulent flows.
Following Schwarz \cite{schwarz-1988}, we hence describe vortex lines as space curves 
$\mathbf{s}(\xi,t)$ of infinitesimal thickness carrying one quantum of
circulation
$\kappa=h/m_4=9.97 \times 10^{-8}~\rm m^2/s$, 
where $h$ is Planck's constant,
$m_4=6.65\times10^{-27}\text{kg}$ is the mass of one helium atom,
$\xi$ is arclength and $t$ is time. 
The equation of motion of the vortex line is 

\begin{equation}
\dot{\mathbf{s}}(\xi,t) = \frac{\partial\s}{\partial t} = \mathbf{v}_{s_{\perp}}
+ \beta \mathbf{s}' \times  \left ( \mathbf{v}_n - \mathbf{v}_s \right )
+\beta' \mathbf{s}' \times \left [ \mathbf{s}' \times \left ( \mathbf{v}_n - \mathbf{v}_s \right ) \right ] \; \; ,
\label{eq: schwarz}
\end{equation}

\noindent
where $\s'=\partial\s/\partial\xi$ is the unit tangent vector at $\s$, 
$\v_n$ and $\v_s$ are the normal fluid and superfluid velocities at $\s$,
and $\beta$, $\beta'$ are temperature and Reynolds
number dependent mutual friction coefficients 
\cite{galantucci2020new}. The $v_{s_{\perp}}$ term indicates 
the projection of superfluid velocity on a plane othogonal to $\s'$. 
The superfluid velocity at a point $\x$ 
induced by the entire vortex configuration $\mathcal{T}$ 
is determined by the Biot-Savart law:
\begin{equation}
\v_s(\x,t) = \frac{\kappa}{4\pi}\oint_{\mathcal{T}}\frac{\mathbf{s} '(\xi,t) \times \left[\x - \mathbf{s} (\xi,t)\right]}{|\x - \mathbf{s}(\xi,t)|^3}d\xi \; \; .
\label{eq: BS}
\vspace{1ex}
\end{equation}
Usually Eqs. \eqref{eq: schwarz} and \eqref{eq: BS}
are always meant to be supplemented by a vortex reconnection algorithm 
\cite{baggaley2012sensitivity}) 
to cope with collisions of vortex lines: in this study, such collisions do not occur. This model of vortex lines is valid under the condition that the discretisation on the lines $\Delta\xi$ is smaller than the average vortex distance $\ell$ and much greater than the vortex core radius $a_0$.

Evolving the vortex lines using 
Eqs. \eqref{eq: schwarz} and \eqref{eq: BS}
gives rise to a {\it one-way model} that neglects the back reaction 
of the vortex lines onto the normal fluid. Accounting
for this back reaction, however, is crucial to understand more accurately the 
interaction between the two fluids. A {\it two-way model}
is obtained by evolving the normal fluid self-consistently 
\cite{galantucci2020new} according to the
Navier-Stokes equation for $\v_n$ modified by the introduction
of the mutual friction force per unit volume $\mathbf{F}_{ns}$:

\begin{equation}
\frac{\partial\v_n}{\partial t} + (\v_n\cdot\nabla)\v_n = -\frac{1}{\rho}\nabla p + \nu_n \nabla^2\v_n + \frac{\mathbf{F}_{ns}}{\rho_n}
\label{eq: NS}
\end{equation} 
\begin{equation}
\mathbf{F}_{ns} = \oint_{\mathcal{T}}\mathbf{f}_{ns}(\mathbf{s})\delta(\x-\mathbf{s})d\xi, \quad \nabla\cdot\v_n = 0
\end{equation}
where $\rho =\rho_n + \rho_s$ is the total helium density, $\rho_n$ and 
$\rho_s$ are respectively the normal and superfluid densities, 
$p$ is the pressure, $\nu_n$ is the kinematic viscosity of the normal fluid
and $\mathbf{f}_{ns}$ is the local friction per unit length \cite{galantucci2015coupled}, defined by
\begin{equation}
 \mathbf{f}_{ns}(\s) = -D\s'\times\left[\s'\times(\dot{\s}-\v_n)\right] - \rho_n \kappa \s'\times(\v_n-\dot{\s}),
\end{equation}
where the coefficient $D$ is 
\begin{equation}
	D = \frac{4\pi\rho_n\nu_n}{\left[\frac{1}{2}-\gamma - \ln\left(\frac{|\v_{n_{\perp}} - \dot{\s}|a_0}{4\nu_n}\right)\right]},
\end{equation}
where here $\v_{n_{\perp}}$ represents the normal fluid velocity lying on a plane orthogonal to $\s'$, and $\gamma=0.5772$ is the Euler-Mascheroni constant. The hydrodynamic model of the normal fluid is valid under the continuum approximation of the Navier-Stokes equations; this approximation is valid as the discretisation of the normal fluid grid $\Delta x$ is larger than the roton-roton mean free path $\lambda_{mfp} = 3\nu_n/v_G$, where $v_G=\sqrt{2k_BT/(\pi \mu)}$ is the roton group velocity and $\mu=0.16m_4$ is the effective mass of a roton. The seperation of these two length scales is in fact of several orders of magnitude $\Delta x/\lambda_{mfp}\sim10^4$.
Methods for fully-coupled dynamics have been used 
in recent studies \cite{galantucci2020new, galantucci2023friction, galantucci2015coupled, kivotides2018superfluid, yui2020fully}: the results presented here are based on \cite{galantucci2023friction,galantucci2015coupled}. The counterflow velocity is forced in the normal fluid by imposing the 0-th Fourier mode in our spectral code, while in the superfluid component by imposing a constant background vector.\\

In this paper, we use the two-way model to solve the two problems described
in Section \ref{intro}. We solve the governing equations and report
input parameters and results using dimensionless units.
We use a characteristic length scale $\tilde{\lambda} = D/L$, where $D^3=(0.1\mathrm{cm})^3$ is the dimensional cube size, $L^3=(2\pi)^3$ is the size of the non-dimensional cubic computational domain and time scale $\tilde{\tau} = \tilde{\lambda}^2\nu_n^0/\nu_n$, where $\nu_n^0$ is the dimensionless viscosity set to properly resolve the small scales of the normal fluid \cite{galantucci2020new}. For this simulation, they are
$\tilde{\lambda}=1.59\times 10^{-2}\cm$, $\nu_n^0 = 0.04$ and $\tilde{\tau}=4.58\times10^{-2}\sec$.

In both problems, we consider a $L^3=(2\pi)^3$ cube with periodic boundary 
conditions at temperature $T=1.9\K$. Four helical vortex lines aligned in the $z$-direction
with amplitude $A=0.1$ and wavenumber $k=5$ are initialised in a chess-board like configuration, such that each vortex (with anti-clockwise rotation) is exactly $L/2$ away from an anti-vortex (with clockwise rotation) and is left to evolve in time. This vortex configuration, unlike a single centralised vortex line, preserves a net-zero circulation under periodic boundary conditions. This one-vortex configuration could lead to inconsistencies due to boundary conditions not being satisfied. However due to the simple nature of the simulation, this effect is not significant.
In the first problem the normal fluid is initially at rest, 
while in the second problem an imposed mean counterflow velocity 
$v_{ns}=1$ in the $z$-direction fuels the Donnelly-Glaberson instability (see Figs. \ref{fig: schematic} and \ref{fig: 2d-schematic}). 
The Lagrangian discretisation along the vortex lines has size
$\delta=0.02$ (corresponding to an initial number of vortex discretisation 
points equal to 1872) with timestep $\Delta t_{_{VF}}=6.25\times10^{-5}$. 
To solve Eq. (\ref{eq: NS}) for the normal fluid we use an
Eulerian discretisation with $N=256^3$ mesh points
 and timestep $\Delta t_{_{NS}} = 40\Delta t_{_{VF}}$. The analysis of the results presented relate to a single vortex line, and global normal fluid quantities are computed across the quadrant in which the vortex resides.  

\section{Results}\label{results}
\subsection{Kelvin wave in initially quiescent normal fluid}\label{subsec:quiescent}

\begin{figure}[t]
	\centering
	\begin{subfigure}[b]{0.49\textwidth}
		\centering
		\includegraphics[width=\textwidth]{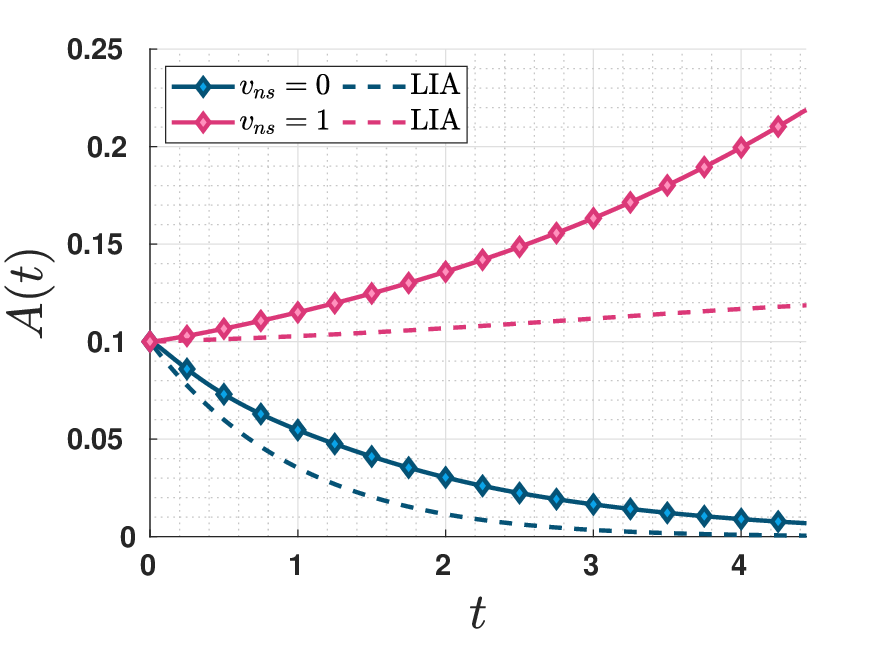}
		\caption{}
		\label{fig: amplitude}
	\end{subfigure}
	\hfill
	\begin{subfigure}[b]{0.49\textwidth}
		\centering
		\includegraphics[width=\textwidth]{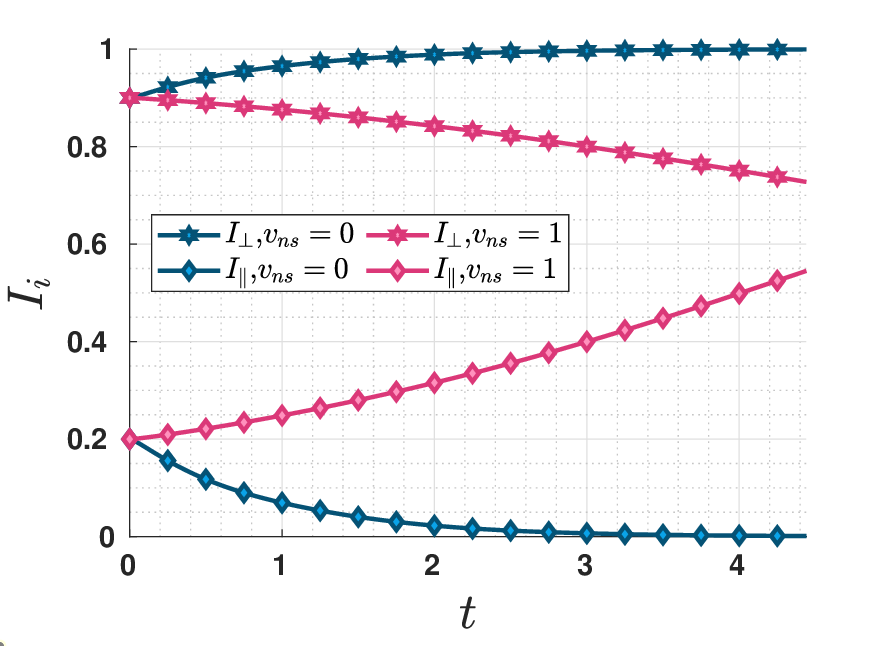}
		\caption{}
		\label{fig: anis-param}
	\end{subfigure}
	
	\caption{\small
\textit{Left:} Evolution of the Kelvin wave's amplitude: comparison between
two-way model (solid lines with red/blue symbols) and one-way model in the LIA (dashed lines).
The blue symbols refer to the first problem (Kelvin wave in initially 
quiescent normal fluid), and the red symbols to the second
problem (Kelvin wave in the presence of counterflow). 
\textit{Right:} Anisotropic parameters $I_{\parallel}$ and $I_{\perp}$ 
computed using the two-way model
corresponding to the first problem (blue symbols) and the
second problem (red symbols). }\label{fig: amplitude&geometry}
\end{figure}

The initial condition of the first problem consists of a
Kelvin wave on a vortex line in an initially quiescent normal fluid. 
At small amplitudes, the Kelvin wave rotates with angular
frequency $\omega \propto k^2$, neglecting logarithmic corrections.
The relative motion between the vortex line and the
normal fluid induces a mutual friction that creates a dipole pattern
in the normal fluid
around the vortex line, as previously reported in the
literature for a single vortex ring
\cite{kivotides-barenghi-samuels-2000}, a single vortex line 
\cite{idowu-willis-barenghi-samuels-2000}) and 
a bundled vortex structure \cite{galantucci2023friction}. In our case
the dipole pattern is twisted in a helical shape,
as illustrated in Fig. \ref{fig: schematic}. 
As the Kelvin wave propagates, 
it injects vorticity and energy into the normal fluid. 
Consequently the superfluid loses 
energy, corresponding to a decrease of the amplitude of the
Kelvin wave. This decrease can clearly be observed
in Fig. \ref{fig: amplitude} (blue curves), where the numerically computed
temporal evolution of the Kelvin wave's amplitude according to the two-way
model is reported along with 
the prediction of the one-way model in the local induction
approximation (LIA), which can be calculated analytically.
We observe that the two-way coupled decay
of the Kelvin wave is slower,  prolonging the lifetime of the Kelvin wave. 
This effect has also been observed in recent numerical
\cite{galantucci2023friction} and experimental \cite{tang2023imaging} 
studies of vortex rings.

The decay of the Kelvin wave towards a straight vortex line can 
also be observed in Fig. \ref{fig: anis-param} (blue curves) 
where we report the temporal evolution of the anisotropic parameters $I_{\parallel}$ and $I_{\perp}$
\cite{schwarz1988three, adachi2010steady}. The parameters quantify the anisotropy of the vortex configuration in the directions parallel and perpendicular to the counterflow direction, which are obtained by

\begin{equation}
I_{\parallel} = \frac{1}{\Omega L'}\oint_{\mathcal{T}}
[1-(\s'\cdot\hat{\r}_{\parallel})^2]d\xi, \label{eq:I.parallel}
\end{equation} 

\begin{equation}
I_{\perp} = \frac{1}{\Omega L'}\oint_{\mathcal{T}}
[1-(\s'\cdot\hat{\r}_{\perp})^2]d\xi,  \label{eq:I.perp}
\end{equation}

where $\hat{\r}_{\parallel}$ is the unit vector parallel to the $z$-direction, $\hat{\r}_{\perp}$ is the unit vector parallel to the $x$-direction,
$L'$ is the vortex line density (vortex length per unit volume)
defined by $L' = (1/\Omega)\oint_{\mathcal{T}}d\xi$, and $\Omega$ 
is the volume of the computational domain. 
The decay of the Kelvin wave into a straight vortex
in fact implies that $I_{\parallel} \rightarrow 0$ and 
$I_{\perp} \rightarrow 1$.

The decaying amplitude $A(t)$ of the Kelvin wave also implies that the magnitude of the velocity difference $\dot{\mathbf{s}} - \v_n$
decreases with time, reducing the magnitude of the mutual friction 
$\mathbf{f}_{ns}$: this effect is reported in 
Fig. \ref{fig: mutual-friction}. 
In the first approximation, in fact, we have
$|\mathbf{f}_{ns}| \propto | \dot{\mathbf{s}} - \v_n | 
\approx |\dot{\mathbf{s}}| \propto \zeta \propto A$, 
where $\zeta=\vert \mathbf{s}''\vert$ is the curvature of
the vortex line at $\mathbf{s}$.

To monitor the transfer of energy between the two fluids, 
we compute the energy injected per unit time 
by the superfluid vortex into the normal fluid, defined as 

\begin{equation}
\epsilon_{inj}=\int_{\Omega} \textbf{F}_{ns}\cdot\v_n d^3\x \;\; , \label{eq:en.inj}
\end{equation}
coinciding with the effective work per unit time performed by the mutual friction force on the normal fluid. This energy injection generates
velocity gradients  ({\it i.e.} vorticity, see Fig. \ref{fig: schematic}) 
into the normal fluid, hence induces the viscous dissipation 

\begin{equation}
\epsilon_{diss}=\nu_n \rho_n \int_{\Omega} \omega^2 d^3\x \;\; ,\label{eq:en.diss}
\end{equation}

The temporal evolutions of $\epsilon_{inj}$ and $\epsilon_{diss}$ 
are reported in Fig. \ref{fig: inj-diss}, where
we observe that the energy injected is rapidly dissipated by viscosity. 
In the initial transient phase, the injection of normal fluid energy dominates, until dissipation takes over.\\

\subsection{Kelvin wave in the presence of counterflow}\label{subsec:DG}

In the second problem, we impose a background counterflow along the
$z$ axis; the average normal fluid velocity 
is in the positive $z$ direction (hence parallel to the superfluid vorticity). 
Provided that $v_n>v_n^c$ where $v_n^c$ is a critical velocity,
the imposed normal flow feeds energy into the superfluid, hence
the amplitude of the Kelvin wave grows
(Donnelly-Glaberson instability). 
The critical velocity $v_n^c$ can be determined
analytically in the one-way model under constant $\v_n$ and the LIA.
The resulting temporal evolution of the amplitude of the Kelvin wave is
$A(t)=A_0 e^{\sigma t}$, where $A_0$ is the initial amplitude,
$\sigma=\alpha k (v_{ns}-\gamma k)$ is the growth rate,
$\gamma=\kappa/(4\pi)\ln(1/(\zeta a_0))$, $\alpha$ is the friction
coefficient in the one-way model,
and $k$ the Kelvin wave's wavenumber. Hence, the Kelvin wave grows 
in amplitude if the amplitude of the imposed
normal fluid is larger than
$v_{n}^c = (\rho_s/\rho)(\kappa/(4\pi)) k \ln(1/(\zeta a_0))$ 
\cite{tsubota2004instability}.
If we choose $v_n > v_n^c$ then the amplitude of the Kelvin wave
grows, as shown in Fig. \ref{fig: amplitude} (red curves). If we compare the growth rate of the two-way model with 
the one obtained analytically with LIA employing the one-fluid model, we obtain a substantial difference, as expected.
In fact, LIA is a too simple framework to describe accurately the evolution of the Kelvin wave, neglecting
the relevance of the logarithmic correction. In addition, LIA is a good approximation only if
$A(t) k \ll 1$ and furthermore
the friction coefficients of the two-way model are different from
the coefficients of the orginal one-way model of Schwarz.

The growth of the Kelvin wave's amplitude with respect to time 
leads tangent vectors to the vortex line to progressively lie 
on a horizontal $xy$ plane. For large times, 
the amplitude of the Kelvin wave becomes comparable to the wavelength,
$A\simeq 2\pi/k$, and the behaviour of the helix becomes similar to
that of stacked vortex rings. At large amplitude, the vortex effectively
lies on the $xy$ plane (orthogonal to the counterflow velocity $\v_{ns}$), 
such that in time $I_{\parallel} \rightarrow 1$ and $I_{\perp} \rightarrow 1/2$, 
as it can be observed in Figs. \ref{fig: anis-param} and \ref{fig: anis-param-extended} (see Appendix \ref{sec: appendix}).

\begin{figure}
	\begin{subfigure}[b]{0.49\textwidth}
		\centering
		\includegraphics[width=\textwidth]{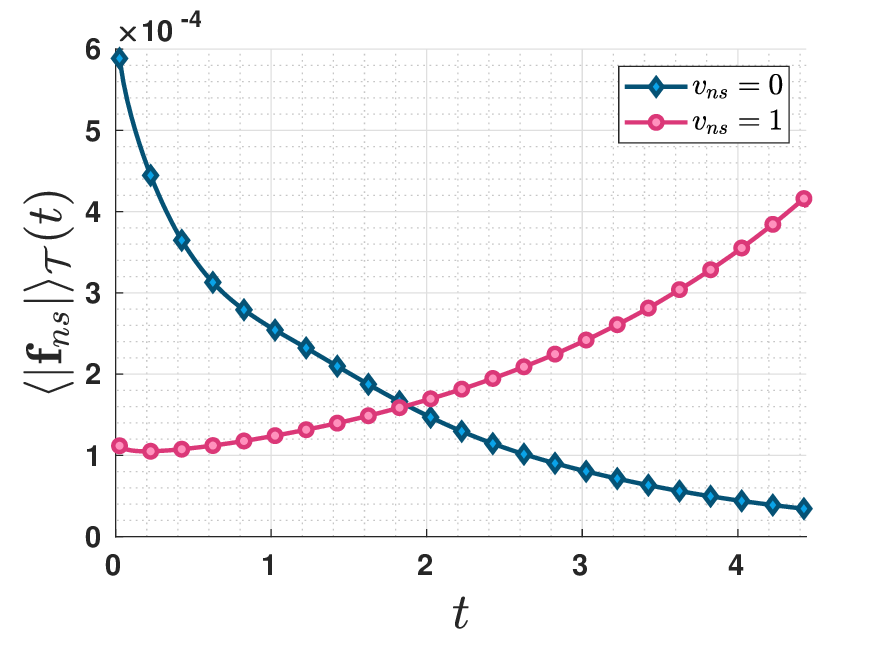}
		\caption{}
		\label{fig: mutual-friction}
	\end{subfigure}
	\hfill
	\begin{subfigure}[b]{0.49\textwidth}
		\centering
		\includegraphics[width=\textwidth]{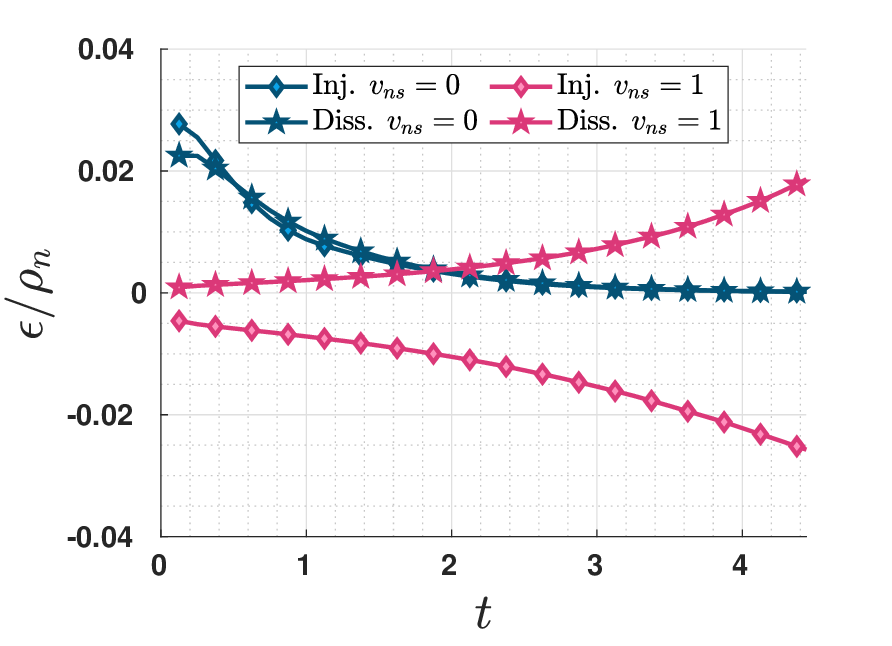}
		\caption{}
		\label{fig: inj-diss}
	\end{subfigure}
	\caption{\small Temporal evolution of the average
mutual friction $\textbf{f}_{ns}(\s)$ (\textit{left}) and of the
injection/dissipation per unit mass of normal fluid
(\textit{right}). The blue symbols refer to the first problem (Kelvin
wave in initially quiescent normal fluid) and the red symbols to the 
second problem Kelvin wave in the presence of counterflow). Note here that unlike the amplitude, $\mathbf{f}_{ns}$ in the quiescent case does not decay to 0, but rather a finite value. The vortices in the configuration do in fact move with a small but non-zero translational velocity induced by the other vortices, stirring the normal fluid.}\label{fig: mutual-friction&normal-fluid}
\end{figure}

The increase of the angle between the counterflow direction and the tangents to the superfluid vortex is responsible for the observed stronger 
intensity of the mutual friction force as time increases, see Fig. \ref{fig: mutual-friction}. The increasing magnitude of $|\mathbf{f}_{ns}|$
leads to an increase of $|\epsilon_{inj}|$ reported in Fig. \ref{fig: inj-diss}. The negative value of $\epsilon_{inj}$ confirms that the energy
is globally transfered from the normal fluid to the superfluid vortex. As in the previous numerical experiment where the normal fluid is initially
quiescent, the motion of the superfluid vortex lines injects vorticity in the normal fluid resulting in the onset of viscous dissipation. It is 
worth noting that compared to the quiescent case, the magnitudes of energy injection and dissipation exhibit a completely different character: 
both quantities do in fact increase with time.\\

\section{Conclusion}\label{conclusions}

We have studied the transfer of energy between normal fluid and
superfluid numerically in two simple problems in which the motion of
the fluids is laminar and the vortex configuration is simple. The main
feature of our study is the use of a recently-developed two-way model
\cite{galantucci2020new} capable of accounting for experimental results 
reported in literature about mildly turbulent bundles of vortex rings
\cite{borner-etal-1983,tang2023imaging}. We have related
the evolution of the geometry of the vortex line to
the normal fluid energy injection and dissipation.
In the first problem (a Kelvin wave propagating in a normal fluid
background initially at rest) energy is transfered from the superfluid
into the normal fluid and eventually dissipated into heat.
In the second problem (a Kelvin wave in the
presence of an imposed thermal counterflow) both energy injection and
dissipation increase with time.
When compared to analytical results obtained using the one-way Local Induction Approximation (LIA), in the counterflow case 
the self-consistent nature of the two-way model drives a significantly 
more rapid amplitude growth, while in the quiescent case, with respect to 
LIA, the Kelvin waves are less attenuated by motions induced in the
normal fluid around the vortex line. The analytic derivation of LIA is based on the amplitude being much smaller in comparison with the wavelength, an approximation which becomes invalid as the amplitude increases. As this occurs, the non-local interactions which are neglected in LIA become significant in the dynamics, and hence LIA is not a valid approximation. As expected, LIA is a too simplified model for the systems investigated in this work. 

These results strengthen our understanding of two-fluid
hydrodynamics, which will be soon tested in on-going experiments 
\cite{peretti-et-al-2023} on decay and amplification of Kelvin waves
in a rotating cryostat.

\section*{Acknowledgments}
AWB acknowledges the support of the Leverhulme Trust through the Research Project Grant RPG-2021-108. GK acknowledges financial support from the Agence Nationale de la Recherche through the project GIANTE ANR-18-CE30-0020-01. CFB acknowledges the support of UKRI grant
Quantum simulators for fundamental physics (ST/T006900/1). LG acknowledges the support of Istituto Nazionale di Alta Matematica (INDAM).

\bibliographystyle{unsrt}

\appendix

\section{Extended anistropic parameters figure}\label{sec: appendix}

\begin{figure}[H]
	\centering
	\includegraphics[width=0.45\textwidth]{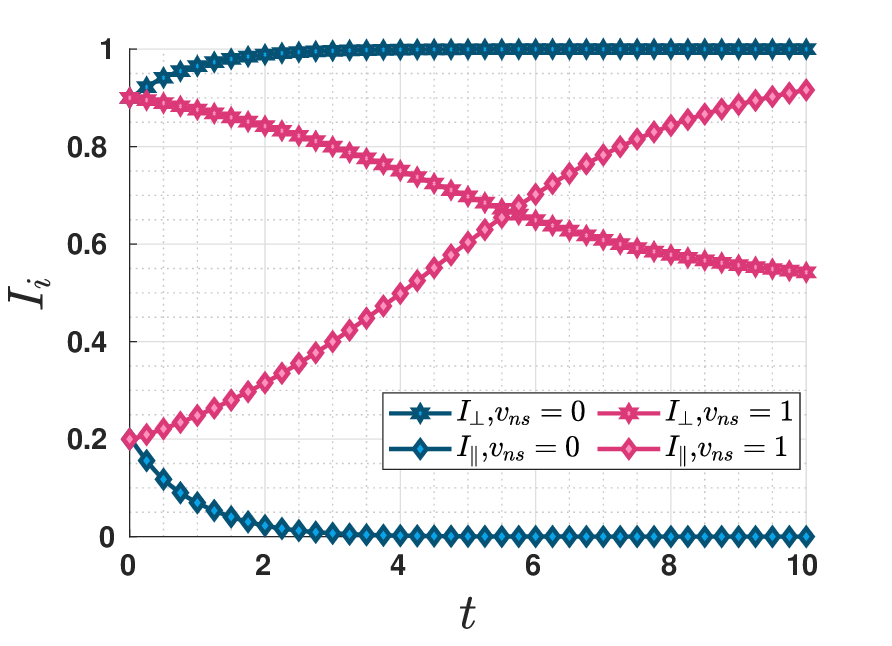}
	\caption{The full time evolution of the parameters $I_{\parallel}$ and $I_{\perp}$.}
\label{fig: anis-param-extended}
\end{figure}

For consistency, the time duration of the plots in Figs. \ref{fig: amplitude&geometry} and \ref{fig: mutual-friction&normal-fluid} was constrained to $t\sim 4.5$, such that the re-meshing \cite{galantucci2020new} of vortex filaments did not factor into the analysis. In particular, the mutual friction in Fig. \ref{fig: mutual-friction} is affected by the re-meshing algorithm, and therefore our analysis is only relevant in the duration up to the re-meshing of the vortex filaments in the counterflow case. Since the anisotropic parameters $I_{\parallel}$ and $I_{\perp}$ are dependent on the geometry of the vortex configuration only, they are not affected by the re-meshing algorithm (see Fig. \ref{fig: anis-param-extended})

In the counterflow case, as the amplitude of the Kelvin wave becomes sufficiently large $Ak>>1$, the non-local interaction dominates the local interaction. The resulting configuration resembles a stack of $k$ vortex rings. A vortex ring of radius $R$ can be parameterised by the arclength $\xi$ to give $\s(\xi) = (R\cos(\xi/R),R\sin(\xi/R),z)$. The parallel anisotropic parameter gives $I_{\parallel}=1$, while the perpendicular gives
\begin{equation}
	I_{\perp} = \frac{1}{\Omega L'}\int_0^{2\pi R}1-(\s'\cdot r_{\perp})^2 d\xi = \frac{1}{2\pi R}\int_0^{2\pi R}\cos^2(\xi/R)d\xi = \frac{1}{2}.
\end{equation}
In the later stages of Fig. \ref{fig: anis-param-extended}, it can be seen that the parameters in the counterflow case begin to asymptote to the values as calculated for a single vortex ring. This confirms the geometry of the vortex configuration, resembling stacked vortex rings.

\end{document}